\documentclass[aps,prl,preprint,showpacs,showkeys,superscriptaddress]{revtex4}

\newcommand{\ts}{\textsuperscript}
\newcommand{\beupl}{\ensuremath{B(E2;~0^+_{\mathrm{gs}}\rightarrow2^+_1)}} 
\newcommand{\beup}{\ensuremath{{B(E2)\!\!\uparrow}}}  
\newcommand{\bevalnu}{241(31)}  
\newcommand{\beval}{\ensuremath{\bevalnu~e^2\mathrm{fm}^4}}

\usepackage{graphicx}
\usepackage{amssymb}

\begin{document}

%
%
\title{``Safe'' Coulomb Excitation of \textsuperscript{30}Mg}

\author{O.~Niedermaier}         
\author{H.~Scheit}              
\email{h.scheit@mpi-hd.mpg.de}
\author{V.~Bildstein}		
\author{H.~Boie}		
\author{J.~Fitting}  		
\author{R.~von~Hahn}		
\author{F.~K\"ock}		
\author{M.~Lauer}		
\author{U.K.~Pal}		
\author{H.~Podlech}             
\author{R.~Repnow}              
\author{D.~Schwalm}		
\affiliation{Max-Planck-Insitut f\"ur Kernphysik, Heidelberg, Germany}
\homepage{http://www.mpi-hd.mpg.de/cb/}

\author{C.~Alvarez}		
\author{F.~Ames}		
\author{G.~Bollen}              
\author{S.~Emhofer}		
\author{D.~Habs}		
\author{O.~Kester}		
\author{R.~Lutter}		
\author{K.~Rudolph}		
\author{M.~Pasini}		
\author{P.G.~Thirolf}		
\author{B.H.~Wolf}		
\affiliation{Ludwig-Maximilians-Universit\"at, M\"unchen, Germany}

\author{J.~Eberth}		
\author{G.~Gersch}		
\author{H.~Hess}		
\author{P.~Reiter}		
\author{O.~Thelen}		
\author{N.~Warr}		
\author{D.~Weisshaar}		
\affiliation{Institut f\"ur Kernphysik, Universit\"at K\"oln, K\"oln, Germany}

\author{F.~Aksouh}		
\author{P.~Van~den~Bergh}       
\author{P.~Van~Duppen}		
\author{M.~Huyse}		
\author{O.~Ivanov}		
\author{P.~Mayet}		
\author{J.~Van~de~Walle}	
\affiliation{Instituut voor Kern- en Stralingsfysica, University of Leuven, Leuven, Belgium}

\author{J.~\"Ayst\"o}           
\author{P.A.~Butler}		
\affiliation{CERN, Geneva, Switzerland} 
\author{J.~Cederk\"all}		
\affiliation{Max-Planck-Insitut f\"ur Kernphysik, Heidelberg, Germany}
\affiliation{CERN, Geneva, Switzerland} 
\author{P.~Delahaye}		
\author{H.O.U.~Fynbo}           
\author{L.M.~Fraile}            
\author{O.~Forstner}            
\affiliation{CERN, Geneva, Switzerland} 
\author{S.~Franchoo}		
\affiliation{CERN, Geneva, Switzerland}
\affiliation{Johannes-Gutenberg-Universit\"at, Mainz, Germany}
\author{U.~K\"oster}		
\affiliation{CERN, Geneva, Switzerland}
\author{T.~Nilsson}             
\affiliation{CERN, Geneva, Switzerland}
\affiliation{Institut f\"ur Kernphysik, Technische Universit\"at Darmstadt, Darmstadt, Germany}
\author{M.~Oinonen}             
\author{T.~Sieber}		
\author{F.~Wenander}		
\affiliation{CERN, Geneva, Switzerland}

\author{M.~Pantea}		
\author{A.~Richter}		
\author{G.~Schrieder}		
\author{H.~Simon}		
\affiliation{Institut f\"ur Kernphysik, Technische Universit\"at Darmstadt, Darmstadt, Germany}

\author{T.~Behrens}		
\author{R.~Gernh\"auser}	
\author{T.~Kr\"oll}		
\author{R.~Kr\"ucken}		
\author{M.~M\"unch}		
\affiliation{Technische Universit\"at M\"unchen, Garching, Germany}

\author{T.~Davinson}		
\affiliation{University of Edinburgh, Edinburgh, United Kingdom}
\author{J.~Gerl}                
\affiliation{Gesellschaft f\"ur Schwerionenforschung, Darmstadt, Germany}
\author{G.~Huber}               
\affiliation{Johannes Gutenberg-Universit\"at, Mainz, Germany}
\author{A.~Hurst}		
\affiliation{Oliver Lodge Laboratory, University of Liverpool, United Kingdom}
\author{J.~Iwanicki}		
\affiliation{Heavy Ion Laboratory, Warsaw University, Warsaw, Poland}
\author{B.~Jonson}              
\affiliation{Chalmers Tekniska H\"ogskola, G\"oteborg, Sweden}
\author{P.~Lieb}                
\affiliation{Georg-August-Universit\"at, G\"ottingen, Germany}
\author{L.~Liljeby}             
\affiliation{Manne Siegbahn Laboratory, Stockholm, Sweden}
\author{A.~Schempp}             
\affiliation{Johann Wolfgang Goethe-Universit\"at, Frankfurt, Germany}
\author{A.~Scherillo}		
\affiliation{Institut f\"ur Kernphysik, Universit\"at K\"oln, K\"oln, Germany}
\affiliation{Institut Laue-Langevin, Grenoble, France}
\author{P.~Schmidt}             
\affiliation{Johannes Gutenberg-Universit\"at, Mainz, Germany}
\author{G.~Walter}              
\affiliation{Institut de Recherches Subatomiques, Strasbourg, France}

\date{\today}
\pacs{25.70.De, 27.30.+t, 21.10.Re}
\keywords{Coulomb excitation, $20\le A\le38$, Collective levels, Radioactive beams}

%
%
\begin{abstract}
We report on the first radioactive beam experiment performed 
at the recently commissioned REX-ISOLDE
facility at CERN in conjunction with the highly efficient $\gamma$ spectrometer
MINIBALL. 
Using \ts{30}Mg ions accelerated to an energy of 2.25 MeV/u 
together with a thin \ts{nat}Ni target, Coulomb excitation of the first excited 
$2^+$ states of the projectile and target nuclei well below the Coulomb barrier
was observed. 
From the measured relative de-excitation $\gamma$ ray yields 
the \beupl\ value of \ts{30}Mg was determined to be \beval.
Our result is lower than values
obtained at projectile 
fragmentation facilities using the intermediate-energy Coulomb excitation method, 
and confirms the theoretical conjecture that the neutron-rich
magnesium isotope \ts{30}Mg lies still outside 
the ``island of inversion''.
\end{abstract}

%
%
\maketitle

%
%
The surprising finding by Thibault \textit{et al.}, that
the neutron-rich sodium isotopes \ts{31}Na and \ts{32}Na are more tightly
bound than expected by sd shell model calculations~\cite{thibault:1975}, 
and its subsequent interpretation by Campi \textit{et al.}~\cite{campi:1975}
has stirred much interest in this region of the nuclear chart.
These nuclei are now considered, together with the neutron-rich Ne
and Mg isotopes, to belong to the
so-called ``island of inversion''~\cite{warburton:1990}, where     
strongly deformed intruder configurations involving neutron 
excitations across a melted $N=20$ shell gap are dominating the 
ground state wave functions. 
However, despite considerable theoretical
and experimental efforts, the question as to where in the $Z$--$N$ plane 
and how rapid the transition from normal to intruder-dominated configurations
takes place is still uncertain;
even the origin of the large collectivity of the $0^+_{\mathrm{gs}}$ to $2^+_1$ 
transition in \ts{32}Mg is still under debate~\cite{yamagami:2004}.

A characteristic feature of isotopes that belong
to the ``island of inversion'' is the presence of
highly collective
$E2$ transitions 
between the low lying states. 
The availability   
of beams of these exotic nuclei at
projectile fragmentation facilities in the 90's 
therefore prompted several of these laboratories   
to start programs to measure \beupl\
values of even-even nuclei in this mass region using the method of intermediate-energy 
Coulomb excitation~\cite{motobayashi:1995}. For the neutron-rich
\ts{30,32,34}Mg isotopes, for example, \beup\ values have been obtained by three groups 
working at RIKEN~\cite{motobayashi:1995,iwasaki:2001}, MSU~\cite{pritychenko:1999}, and 
GANIL~\cite{chiste:2001} (see also Fig.~\ref{fig:mgbe2}), however,  
their results are neither conclusive nor consistent.  
Differences between the data points for a given isotope are as large as a factor of two
and do not allow the various theoretical
model predictions~\cite{utsuno:1999,caurier:2001,rodriguez-guzman:2002}
to be distinguished,
nor can any firm conclusion be drawn from these data alone regarding the
boundary of the ``island of inversion''.

In an attempt to clarify the experimental situation, with the long term goal to
map the collectivity of nuclei in and around the ``island of inversion'',
an experimental program was started to measure
the \beup\ values with a standard 
model-independent technique, namely ``safe'' Coulomb excitation
in reversed kinematics, employing ISOL beams accelerated to energies well below
the Coulomb barrier in conjunction with a high-resolution detector system.
We report here on the first experiment performed with the 
isotope \ts{30}Mg 
using the newly commissioned REX-ISOLDE 
accelerator~\cite{kester:2003}, located at the ISOLDE 
facility~\cite{kugler:2000} at CERN, together with the high-resolution 
$\gamma$ detector array MINIBALL~\cite{eberth:2001} and 
ancillary Si detectors~\cite{ostrowski:2002}.

%
%
The radioactive \ts{30}Mg atoms ($t_{\frac 12} = 335(17)$~ms) 
were produced by sending 1.4 GeV protons, provided by
the CERN PS Booster with a maximum intensity of 3.2$\cdot10^{13}$~p/pulse 
and a repetition time of typically 1.2 or 2.4 s,  
onto an uranium carbide/graphite target. 
The Mg atoms diffusing out of the target were selectively ionized in the Resonance 
Ionization Laser Ion Source~\cite{fedoseyev:2000} and
the extracted $1^+$ ions were mass separated by the ISOLDE General Purpose Separator~\cite{kugler:2000}.

The REX-ISOLDE accelerator~\cite{kester:2003}, which employs novel techniques to 
accumulate, bunch, charge breed, and accelerate radioactive ions,
was used to boost the energy of the \ts{30}Mg ions to 2.25 MeV/u. 
The $1^+$ ions delivered by ISOLDE were first accumulated, cooled, 
and bunched in a Penning trap for up to 20~ms before they were 
transfered to an electron beam ion-source (EBIS), 
where they were ionized within 12~ms to a charge state of $7^+$. 
The highly charged ions were then extracted in $\sim\! 50$--$100~\mu$s long 
pulses, mass separated with a $q/A$ resolution of 100, and finally injected into a linear
accelerator consisting of an RFQ, an IH-structure, and three 7-gap resonators.
The cycle frequency for this process was 49 Hz.
The average intensity of the \ts{30}Mg\ts{7+} beam on the secondary target was 
about $2\cdot10^4$ s\ts{-1} with an efficiency of the REX accelerator (including trap and EBIS)
of about 5\%.

The \ts{30}Mg ions were incident on a natural nickel foil 
of 1.0 mg/cm$^2$ located in the center of a small scattering chamber.  
Scattered projectiles and recoiling target nuclei were detected by a 500 $\mu$m thick, 
compact-disk-shaped double sided silicon strip detector (CD)~\cite{ostrowski:2002}, 
which is subdivided into four independent quadrants with 
24 sector strips and 16 annular strips each.
The detector covered forward angles between
16.4$^\circ$ and 53.3$^\circ$ degrees. 
Scattered projectiles
and recoiling target nuclei could be well separated via their different energies
at a given laboratory angle. 

The de-excitation $\gamma$ rays following the Coulomb excitation of the  
projectile and target nuclei 
were detected with the MINIBALL array~\cite{eberth:2001},
consisting of eight triple cluster detectors, each combining three six-fold segmented
HPGe detectors. 
By choosing a target-detector distance of only 9~cm,
the cluster detectors covered laboratory angles from $30^\circ$ to $85^\circ$ 
and $95^\circ$ to $150^\circ$,
and an overall full-energy peak efficiency of about 7\% 
at $E_\gamma = 1.3$~MeV (with cluster add-back) could be achieved. 
The interaction point of each $\gamma$ ray was determined by an
online-onboard pulse shape analysis~\cite{lauer:2004}, 
resulting in an about 100-fold increase in granularity in comparison to an array of
non-segmented HPGe detectors. 
Together with the direction of the projectile or target nucleus 
measured in coincidence it was 
therefore possible to correct for the large Doppler shifts 
of the $\gamma$ rays, which are caused by the high velocities 
of the $\gamma$-emitting nuclei ($v/c \sim 5\%$).

\begin{figure}
\includegraphics[width=5.2cm,angle=-90]{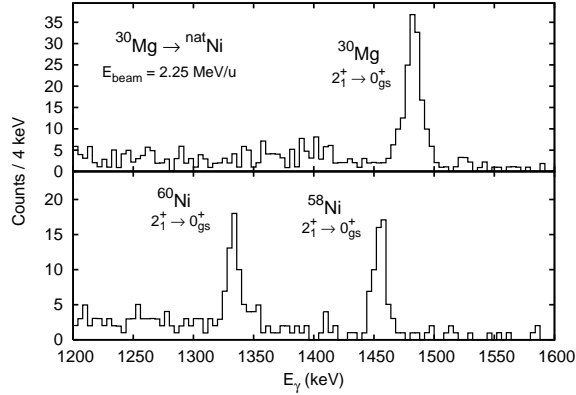}
\caption{\label{fig:spec_MgCD_MgNiDC} Doppler-corrected $\gamma$-ray 
spectra observed in coincidence with projectiles in the CD detector. 
The upper panel shows the spectrum performing a
Doppler correction relevant for $\gamma$ emission from the detected projectile
(without Ni-line contributions, see text), 
while the Doppler-corrected spectrum assuming 
$\gamma$ emission from the corresponding recoiling target nucleus 
is displayed in the lower panel (without the Mg-line contributions, see text).}
\end{figure}

%
%
The $\gamma$-ray energy spectra observed after 76 hours of data taking  
in coincidence with projectiles in the CD detector  
are shown in Fig.~\ref{fig:spec_MgCD_MgNiDC}, with the
upper and lower panel displaying part of the Doppler-corrected 
spectrum assuming the $\gamma$ emitting
nucleus to be the projectile or the recoil, respectively.
The prominent peak in the upper panel observed at an energy of 
1482 keV corresponds to 
the transition from the first $2^+$ state to the ground state of \ts{30}Mg,
while the two lines observed at 1454 keV and 1333 keV in the bottom spectrum
result from the decay of the 
first excited $2^+$ state in \ts{58}Ni and \ts{60}Ni, respectively.
Note that Doppler-smeared contributions of the Ni lines in the upper
spectrum were avoided by suppressing events contributing to the 
Ni lines in the lower spectrum, and vice versa. 
The influence of this procedure
on the line intensities was carefully investigated and resulted only in small 
corrections to the deduced intensity values.

%
%
In order to fulfill the ``safe'' Coulomb excitation condition~\cite{schwalm:1994}, 
i.e. to ensure that only the electromagnetic interaction is causing the excitation
of the projectile and target nuclei, 
the analysis of the observed $\gamma$ intensities in terms of \beup\ values 
was restricted to events observed in coincidence 
with forward scattered \ts{30}Mg,
for which the closest distance   
$D_{\mathrm{s}}$ between the surfaces of the projectile and target nucleus does
not drop below 6~fm~\cite{schwalm:1994}.
Due to the occurrence of both projectile and target excitation the Coulomb excitation
cross section $\sigma_{\mathrm{CE}}$ to the first excited $2^+$ state in \ts{30}Mg 
can be deduced 
relative to that of \ts{58,60}Ni 
from the measured $\gamma$-ray yields $N_\gamma$ by
\[
\frac{\sigma_{\mathrm{CE}}\left(^{30}\mathrm{Mg}\right)}
   {\sigma_{\mathrm{CE}}\left(^{58,60}\mathrm{Ni}\right)} = 
   \frac{\epsilon_\gamma(^{58,60}\mathrm{Ni})}{\epsilon_\gamma(^{30}\mathrm{Mg})}
   \cdot \frac{W_\gamma(^{58,60}\mathrm{Ni})}{W_\gamma(^{30}\mathrm{Mg})}
   \cdot \frac{N_\gamma(^{30}\mathrm{Mg})}{N_\gamma(^{58,60}\mathrm{Ni})},
\]
where $\epsilon_\gamma$ is the full energy peak efficiency at the 
corresponding $\gamma$ energy and $W_\gamma$ the angular correlation factor for the
respective transitions. The Coulomb excitation cross sections $\sigma_{\mathrm{CE}}$,
which are in first approximation proportional to the corresponding \beup\ value,
as well as the angular correlation factors $W_\gamma$ 
were calculated using a standard multiple Coulomb excitation code~\cite{clx},
taking into account the energy loss of the beam in the target and the angles subtended 
by the CD detector and the MINIBALL array. For the Ni isotopes known values for
the \beup\ and quadrupole moments were used~\cite{king:1993,bhat:1997}, while 
for Mg the \beup\ value was varied (assuming a Q(2$^+$) moment as 
expected within the rotational model for a prolate deformed nucleus --- 
the assumption of an oblate deformation would reduce the extracted \beup\ by 12\%)
until the experimental value was reproduced.  
The analysis was performed separately for the two nickel isotopes and the weighted
average of the results was taken as the final value. 

Even though the applied procedure is straightforward, a measurement with a stable
\ts{22}Ne beam of 2.25 MeV/u was performed for test 
purposes~\cite{niedermaier:2004long}; the deduced 
\beup\ of $243(27)~e^2\mathrm{fm}^4$ was not only found to be
in excellent agreement with the literature
value~\cite{raman:2001} of $230(10)~e^2\mathrm{fm}^4$, but it also confirmed our choice
to neglect the recently published~\cite{kenn:2001} \beup\ values for \ts{58,60}Ni, 
which are in contradiction to all previous measurements~\cite{king:1993,bhat:1997}.

%
In contrast to relative Coulomb excitation experiments with stable isotopes, 
however, in measurements with  
radioactive ions possible beam contaminations have to be carefully investigated.  
While contaminations with $A \ne 30$ can be excluded in the present study 
due to the $q/A$ selection of the REX separator and  
the measurement of the total projectile energy in the CD detector,
there are in principle three sources for isobaric contaminations: 
$\beta$ decay products of \ts{30}Mg collected during trapping and charge breeding,
isobaric contaminants directly released from and ionized at the primary ISOLDE target,
and residual gas contaminations produced in the EBIS source. 
The first contribution, which is inherent to our technique, 
can be estimated from the trapping and breeding time,
which ranged from 12~ms to 32~ms depending on the time the \ts{30}Mg ion 
entered the trap; based on the known lifetime of \ts{30}Mg of 483(25) ms 
an \ts{30}Al contamination of the beam of 4.5(0.5)~\% is calculated.
Possible isobaric beam contributions from the second source, which are expected to 
mainly consist of \ts{30}Al as other isobars
have negligible yields
and from the EBIS residual gas were 
investigated by following means:
(i)~A LASER-on/off measurement was performed.
(ii)~The time dependence of the incident beam intensity
with respect to the proton pulse impact on the ISOLDE target (T1) was analyzed;
the \ts{30}Mg ions
show a high intensity only for short times after proton impact due to their 
fast release and short lifetime. 
(iii)~The time dependence of the $\gamma$ yields for \ts{30}Mg and
\ts{58,60}Ni with respect to T1 was studied. 
(iv)~The Coulomb excitation of the first excited state in \ts{30}Al at 244 keV
was searched for. 
(v)~The $\gamma$ intensities due to 
$\beta$ decay of \ts{30}Mg and \ts{30}Al
collected in the target chamber were analyzed. 
The various investigations resulted in a consistent picture for the purity of the beam, 
in that the only noticeable contamination is \ts{30}Al. 
The combined analysis yielded a total \ts{30}Al contribution
of 6.5(1.0)~\% to the \ts{30}Mg beam within the window of $t-t_{\mathrm{T1}}\le1.2$~s,
which was applied when extracting the $\gamma$ intensities. 
This contribution leads to corrections of the measured $\gamma$ intensities of the Ni
transitions of $-5.0(1.0)$~\%.

%
Including the correction due to the \ts{30}Al beam impurity, a \beupl\ value
of \beval\ was determined for \ts{30}Mg. 
The quoted one $\sigma$ error is dominated by the statistical error, 
but also includes those  
caused by the $E2$ matrix elements adopted for \ts{58,60}Ni, 
by the assumed Q(2$^+$) value for \ts{30}Mg and 
possible excitations to higher lying states,
and takes care of uncertainties in the correlation functions $W_\gamma$ caused by
possible deorientation effects~\cite{niedermaier:2004long}.

\begin{figure}
\includegraphics[width=6.0cm,angle=-90]{mgbe2.epsi}
\caption{\label{fig:mgbe2} Experimental (open and filled circles) 
and theoretical \beup\ values (connected by thin lines to guide the eye) 
for the even Mg isotopes. The experimental data are from
Refs.~\cite{motobayashi:1995,iwasaki:2001,pritychenko:1999,chiste:2001,raman:2001} 
and the present experiment, the theoretical values are from:
$\square$~\cite{rodriguez-guzman:2002},
$\lozenge$~\cite{utsuno:1999},
$\triangledown$~\cite{caurier:1998,caurier:2001} (normal),
$\vartriangle$~\cite{caurier:2001} (intruder).
}
\end{figure}

%
%
The present \beup\ value for \ts{30}Mg of \beval\, 
has to be compared to the results obtained by the 
MSU and the GANIL groups using the method of intermediate-energy Coulomb 
excitation, which resulted in values of 
295(26)~\ensuremath{e^2\mathrm{fm}^4}~\cite{pritychenko:1999} and
435(58)~\ensuremath{e^2\mathrm{fm}^4}~\cite{chiste:2001}, respectively 
(see also Fig.~\ref{fig:mgbe2}).
While the MSU value is about 20~\% larger but still consistent within errors
with the present value, the GANIL result exceeds our value by 80~\%. 
The origin of the discrepancy is unclear.

In searching for possible sources for these deviations it should be noted that
in intermediate-energy 
measurements at beam energies around 30-50~MeV/u several effects can influence the deduced
\beup\ values such as feeding from higher lying states and 
Coulomb-nuclear interference, which have to be corrected for. 
While the adiabatic cutoff limits
the single-step excitation energy to values below 1-2~MeV in sub-barrier experiments,
as presented here, 
in measurements with intermediate energy beams $2^+$ (or even $1^-$)
states up to 5--10~MeV
excitation energy can be populated~\cite{alder:1975}, which
may feed the first $2^+$ state. 
Unless these feedings are taken into account,
they will likely result in increased \beup\ values.
Based on the apparent absence of feeding transitions in their 
$\gamma$ spectra, no feeding correction was applied to the MSU value~\cite{pritychenko:1999},
while a 15~\% correction deduced from model calculations including Coulomb-nuclear
interferences was applied to the GANIL value~\cite{chiste:2001}. However,
while feeding and interference effects may account for the slightly larger
MSU value as compared to the present one, it is questionable if
uncertainties in their estimate can be the
cause for the large \beup\ value published by the GANIL group.  
The present result, on the other hand, which 
is based on the well established
technique of Coulomb excitation with beam energies well
below the Coulomb barrier, is safe with regard to nuclear
interference effects and is barely influenced by real or virtual excitations
of higher lying states; moreover, due to the 
relative measurement of projectile to target excitation 
it is rather insensitive to systematic experimental uncertainties. 

Our present experimental knowledge of the \beup\ values for the even Mg 
isotopes with $N \ge 12$ is displayed in Fig.~\ref{fig:mgbe2} together with
theoretical predictions obtained within three different model 
approaches~\cite{utsuno:1999,caurier:2001,rodriguez-guzman:2002},
chosen representatively from a large number of recent publications
(see~\cite{yamagami:2004} and references therein). 
It is obvious that precise
values are needed to judge the quality and predictive power of these
calculations. 
In particular, the results by Caurier \textit{et al.}~\cite{caurier:2001}
are interesting as they give the \beup\ values separately for 
the pure ``intruder'' and ``normal'' configurations, 
i.e. with and without excitations across the sd-pf-shell
boundary, respectively, since their calculation did 
not allow the determination of the amount of mixing between
these two configurations.
Our result provides clear evidence that the lowest $0^+$ and $2^+$ state of
the $N=18$ isotope \ts{30}Mg
can still be well described within the sd shell, in agreement with most theoretical
predictions.

In summary, we have presented the result of the first Coulomb excitation experiment performed with 
the newly commissioned REX accelerator and the MINIBALL array, 
which shows the strength of this novel facility for the study of 
nuclei far from stability by
well established, yet adapted, nuclear physics techniques.
The \beupl\ of \ts{30}Mg was measured to be \beval\ 
(8.7~W.u.\ for the corresponding $2^+_1\rightarrow0^+_{\mathrm{gs}}$ transition),
which is lower than those extracted in previous measurements 
performed at intermediate energies.
It supports the theoretical conjecture 
that \ts{30}Mg is still located outside the ``island of inversion''.
In the near future it is planned to extend these studies to other even-even isotopes
in this mass region, in particular to \ts{32}Mg, where only conflicting results 
from intermediate-energy Coulomb excitation experiments are available so far, 
which do not allow to distinguish between the various theoretical predictions.   

\begin{acknowledgments}
Support by
the German BMBF ({\small 06~OK~958, 06~K~167}),
the Belgian FWO-Vlaanderen and IAP,
the UK~EPSRC, and 
the European Commission ({\small TMR~ERBFMRX~CT97-0123, HPRI-CT-1999-00018, HRPI-CT-2001-50033})
is acknowledged
as well as the support by the ISOLDE collaboration.

\end{acknowledgments}

\bibliography{cb_a1,local}
\end{document}